\documentclass[sigconf, nonacm]{acmart}

\usepackage{algorithm,algpseudocode}
\usepackage{array,multirow}
\usepackage{xspace}
\usepackage{pifont}
\usepackage{adjustbox}
\usepackage{xcolor}
\usepackage{colortbl}
\usepackage{pbox}
\usepackage{graphicx}
\usepackage{arydshln}
\usepackage{subfigure}
\usepackage{bm}
\usepackage{amsmath}
\usepackage{makecell}
\usepackage{url}

\newcommand\vldbpagestyle{plain}

\newcommand{\CE}{\textsf{CardEst}\xspace}
\newcommand{\Card}{\textsf{Card}\xspace}

\mathchardef\mhyphen="2D
\newcommand\revise[1]{\textcolor{black}{#1}}

\newcommand{\myskip}{\vspace{0.5em}}
\definecolor{mygrey}{RGB}{230,230,240}

\newcommand{\CEend}{\texttt{CardEst}}

\begin{document}
	\title{Cardinality Estimation in DBMS: A \\ Comprehensive Benchmark Evaluation}
	\author{Yuxing Han$^{1, \#}$, Ziniu Wu$^{1, 2, \#}$, Peizhi Wu$^{3}$, Rong Zhu$^{1, *}$, \\ Jingyi Yang$^{3}$,  Liang Wei Tan$^{3}$, $\!$ Kai Zeng$^{1}$, Gao Cong$^{3}$, Yanzhao Qin$^{1, 4}$, \\Andreas Pfadler$^{1}$, Zhengping Qian$^{1}$, Jingren Zhou$^{1}$, Jiangneng Li$^{1, 3}$, Bin Cui$^{4}$}
	\affiliation{%
	\institution{\LARGE{\textit{$^1$Alibaba Group, $^2$MIT, $^3$Nanyang Technological University, $^4$Peking University}} \\
	\textsf{$^1$red.zr@alibaba-inc.com, $^2$ziniuw@mit.edu, ~~~$^3$gaocong@ntu.edu.sg,~~~$^4$bin.cui@pku.edu.cn}
	\country{}
	}
	\vspace{0.5em}
	}

\allowdisplaybreaks

\begin{abstract}
Cardinality estimation (\CE) plays a significant role in generating high-quality query plans for a query optimizer in DBMS.
In the last decade, an increasing number of advanced \CE methods (especially ML-based) have been proposed with outstanding estimation accuracy and inference latency. 
However, there exists no study that systematically evaluates the quality of these methods and answer the fundamental problem: \textit{to what extent can these methods improve the performance of query optimizer in real-world settings}, which is the ultimate goal of a \CE method.

In this paper, we comprehensively and systematically compare the effectiveness of \CE methods in a real DBMS.
We establish a new benchmark for \CE, which contains a new complex real-world dataset STATS and a diverse query workload STATS-CEB. 
We integrate multiple most representative \CE methods into an open-source database system PostgreSQL, and comprehensively evaluate their true effectiveness in improving query plan quality, and other important aspects affecting their applicability, ranging from inference latency, model size, and training time, to update efficiency and accuracy. We obtain a number of key findings for the \CE methods, under different data and query settings. 
Furthermore, we find that the widely used estimation accuracy metric (Q-Error) cannot distinguish the importance of different sub-plan queries during query optimization and thus cannot truly reflect the \revise{query plan quality generated by \CE methods.} Therefore, we propose a new metric P-Error to evaluate the performance of \CE methods, which overcomes the limitation of Q-Error and is able to reflect the overall end-to-end performance of \CE methods. We have made all of the benchmark data and evaluation code publicly available at \textsf{https://github.com/Nathaniel-Han/End-to-End-CardEst-Benchmark}.

\end{abstract}

\maketitle

\pagestyle{\vldbpagestyle}

\begingroup
\renewcommand\thefootnote{}\footnote{\noindent
	$\#$ The first two authors contribute equally to this paper. \\
	$*$ Corresponding author.
}
\addtocounter{footnote}{-1}
\endgroup

\section{Introduction}
The query optimizer is an integral component in modern DBMSs. It is responsible for generating high-quality execution plans for the input SQL queries. \emph{Cardinality estimation(\CE)} plays a significant role in query optimization. It aims at estimating the result size of all sub-plans of each query and guiding the optimizer to select the optimal join operations. The performance of \CE has a 
critical impact on the quality of the generated query plans.

\myskip

\noindent{\underline{\textbf{Background:}}}
Due to its important role in DBMS, \CE has been extensively studied, by both academic and industrial communities. \revise{Current open-source and commercial \revise{DBMSs} mainly use two traditional \CE methods, namely histogram~\cite{selinger1979access,gunopulos2005selectivity,bruno2001stholes,muralikrishna1988equi,wang2003multi,deshpande2001independence} in PostgreSQL\cite{psql2020} and SQL Server\cite{sqlserver2019} and sampling~\cite{leis2017cardinality,heimel2015self,kiefer2017estimating,zhao2018random, li2016wander} in MySQL~\cite{mysql2020} and MariaDB~\cite{mdb2020}.}
The core task of \CE is to build a compact model capturing data and/or query information. With the prosperity of machine learning (ML), we witness a proliferation of learned methods for \CE in the last three years~\cite{kipf2018learned,hilprecht2019deepdb, sun2019end, yang2019deep,yang2020neurocard,wu2020bayescard,zhu2020flat,hasan2020,wu2021uae,dutt2019selectivity}. These methods could be categorized into two classes, namely query-driven and data-driven.
\revise{Query-driven \CE methods~\cite{kipf2018learned, dutt2019selectivity} build discriminative models mapping featurized queries to their cardinalities while data-driven \CE methods~\cite{yang2019deep,yang2020neurocard, tzoumas2011lightweight, getoor2001selectivity,wu2020bayescard, hilprecht2019deepdb, zhu2020flat} directly model the joint distribution of all attributes.} In comparison with the traditional methods, their estimation accuracy stands out as their models are more sophisticated and fine-grained~\cite{zhu2020flat, yang2019deep, wu2020bayescard, hilprecht2019deepdb}. 

\myskip

\noindent{\underline{\textbf{Motivation:}}}
Despite the recent advance of the \CE methods, we notice that a fundamental problem has not yet been answered, which is \emph{``to what extent can these advanced \CE methods improve the performance of query optimizers in real-world settings?''} Although existing studies have conducted extensive experiments, they suffer from the following shortcomings: 

\indent 1. \emph{The data and query workloads used for evaluation may not well represent the real-world scenarios.}
The widely adopted JOB-LIGHT query workload on IMDB benchmark data~\cite{leis2015good} touches at most 8 numerical or categorical attributes within six tables, whose schema forms a star-join. The recent benchmark work~\cite{wang2020ready} only evaluate these methods in a single table scenario. Therefore, the existing works are not sufficient to reflect the behavior of \CE methods on complex real-world data with high skewness and correlations and multi-table queries with various join forms and conditions.

\indent 2. \emph{Most of the evaluations do not exhibit the end-to-end improvement of \CE methods on the query optimizer.}
Existing works usually evaluate \CE methods on the algorithm-level metrics, such as estimation accuracy and inference latency. \revise{These metrics only evaluate the quality of the \CE algorithm itself, but cannot reflect how these methods behave in a real DBMS due to two reasons. First, the estimation accuracy does not directly equal to the query plan quality. As different sub-plan queries matters differently to the query plan~\cite{perron2019learned,chaudhuri2009exact,trummer2019exact}, a more accurate method may produce a much worse query plan if they mistake a few very important estimations~\cite{negi2020cost}.
Second, the actual query time is affected by multiple factors, including both query plan quality and \CE inference cost. 
}
Therefore, the ``gold standard'' to examine a \CE method is to integrate it into the query optimizer of a real DBMS and record the \emph{end-to-end query time}, including both query plan generation time and execution time. Unfortunately, this end-to-end evaluation has been ignored in most existing works.

To address these two problems, the DBMS community needs 
1) \emph{new benchmark datasets and query workloads that can represent the real-world settings} and 2) \emph{an in-depth end-to-end evaluation to analyze performance of \CE methods}. 

\smallskip
\noindent{\underline{\textbf{Contributions and Findings:}}}
In this paper, we provide a systematic evaluation on representative \CE methods and make the following contributions:

\indent 1. \emph{We establish a new benchmark for \CE that can represent real-world settings.}
Our benchmark includes a real-world dataset STATS and a hand-picked query workload STATS-CEB. STATS has complex properties such as large attribute numbers, strong distribution skewness, high attribute correlations, and complicated join schema. STATS-CEB contains a number of diverse multi-table join queries varying in the number of involved tables, true cardinality, and different join types (e.g., chain/star/mixed, one-to-many/many-to-many, etc.).
This benchmark pose challenges to better reveal the advantages and drawbacks of existing \CE methods in real-world settings. (in Section~3) 

\indent \emph{2. We provide an end-to-end evaluation platform for \CE and present a comprehensive evaluation and analysis of the representative \CE methods.}
We provide an approach that could integrate any \CE method in the built-in query optimizer of PostgreSQL, a well-known open-source DBMS. Based on this, we evaluate the performance of both traditional and ML-based \CE methods in terms of the end-to-end query time and other important aspects affecting their applicability, including inference latency, model size, training time, update efficiency, and update accuracy. From the results, we make a dozen of key observations (O). Some key take-away findings are listed as follows (in Sections~4--6):

\indent \textbf{K1. Improvement (O1):}
On numerical and categorical query workloads, the ML-based data-driven \CE methods can achieve \revise{remarkable} performance, whereas \revise{most of} the other methods can hardly improve the \textsf{PostgreSQL} baseline.

\indent \textbf{K2. Method (O3, O8-10):}
\revise{Among the data-driven methods, probabilistic graphical models outperform deep models in terms of both end-to-end query time and other practicality aspects and are more applicable to deploy in real-world DBMS.}

%


\indent \textbf{K3. Accuracy (O5-6, O11-13):}
\revise{
Accurate estimation of some important queries, e.g., with large cardinality, is crucial to the overall query performance. The widely used accuracy metric Q-Error~\cite{moerkotte2009preventing} cannot reflect a method's end-to-end query performance.
}

\indent \textbf{K4. Latency (O7):}
\revise{The inference latency of \CE methods has a non-trivial impact on the end-to-end query time on OLTP workload.
}

\indent \emph{3. We propose	a new metric that can indicate the overall quality of \CE methods.}
Previous \CE quality metrics, such as Q-Error, can only reflect the estimation accuracy of each (sub-plan) query but not the overall end-to-end performance of \CE methods. 
Therefore, inspired by the recent work~\cite{negi2020cost,negi2021flow}, we propose a new metric called P-Error, which directly relates the estimation accuracy of (sub-plan) queries to the ultimate query execution plan quality of the \CE methods.
Based on our analysis, 
P-Error is highly correlated with the end-to-end query time improvement. Thus, it could serve as a potential substitute for Q-Error and a better optimization objective for learned \CE methods. (in Section~7)

\indent 
\revise{
\emph{4. We point out some future research directions for \CE methods.}
On the application scope, future ML-based \CE methods should enhance the ML models to support more types of queries. Moreover, it is also helpful to unify different approaches  and/or models to adjust \CE for different setting, i.e., OLTP and OLAP. On designing principles, we should optimize ML models towards the end-to-end performance metrics instead of purely accuracy metrics, with an emphasis on multi-table join queries.
(in Section~8)
}

\section{Preliminaries and Background}
\label{sec: prelim}

In this section, we introduce some preliminaries and background, including a formal definition of the cardinality estimation (\CE) problem,
a brief review on representative \CE algorithms and a short analysis on existing \CE benchmarks.

\myskip
\noindent \underline{\textbf{\CE Problem:}}
In the literature, \CE is usually defined as a statistical problem. Let $T$ be a table with $k$ attributes $A = \{A_1, A_2, \dots, A_k \}$. $T$ could either represent a single relational table or a joined table.
In this paper, we assume each attribute $A_i$ for each $1 \leq i \leq k$ to be either categorical (whose values can be mapped to integers) or continuous, whose domain (all unique values) is denoted as $D_i$. 
Thereafter, any selection query $Q$ on $T$ can be represented in a canonical form: $Q = \{A_1 \in R_1 \wedge A_2 \in R_2 \wedge \cdots \wedge A_n \in R_n\}$, where $R_i \subseteq D_i$ is the constraint region specified by $Q$ over attribute $A_i$ (i.e. filter predicates). Without loss of generality, we have $R_i = D_i$ if $Q$ has no constraint on $A_i$. Let $\Card(T, Q)$ denote the \emph{cardinality}, i.e., the exact number of records in $T$ satisfying all constraints in $Q$. The \CE problem requires estimating $\Card(T, Q)$ as accurately as possible without executing $Q$ on $T$. 

\revise{
In this paper, we concentrate on evaluating these selection queries on numerical/categorical (n./c.) attributes. We do not consider `LIKE'' (or pattern matching) queries on string attributes due to two reasons: 
1) commercial \CE methods for ``LIKE''  queries in DBMS often use magic numbers~\cite{psql2020, sqlserver2019, mysql2020, mdb2020}, which are not meaningful to evaluate;
and 2) \CE solutions for n./c. queries mainly consider how to build statistical models summarizing attribute and/or query distribution information. Whereas, \CE methods for ``LIKE'' queries ~\cite{sun2019end, shetiya2020astrid, mikolov2013distributed} focus on applying NLP techniques to summarize semantic information in strings. Thus, they tackle with different technical challenges and statistical \CE methods can not effectively support ``LIKE'' queries.
}

\myskip
\noindent \underline{\textbf{\CE Algorithms:}}
There exist many \CE methods in the literature, which can be classified into three classes as follows:

\emph{Traditional \CE methods}, such as histogram~\cite{selinger1979access} and sampling~\cite{leis2017cardinality,heimel2015self,kiefer2017estimating}, are widely applied in DBMS and generally based on simplified assumptions and expert-designed heuristics. 
\revise{
Many variants are proposed later to enhance their performance.
Histogram-based variants include multi-dimensional histogram based methods~\cite{poosala1997selectivity, deshpande2001independence, gunopulos2000approximating, gunopulos2005selectivity, muralikrishna1988equi, wang2003multi}, correcting and self-tuning histograms with query feedbacks~\cite{bruno2001stholes, srivastava2006isomer, khachatryan2015improving, fuchs2007compressed} and updating statistical summaries in DBMS~\cite{stillger2001leo, wu2018towards}.
Sampling-based variants include query-driven kernel-based methods~\cite{heimel2015self,kiefer2017estimating}, index based  methods~\cite{leis2017cardinality} and random walk based  methods~\cite{zhao2018random, li2016wander}.
Some other work, such as the sketch based method~\cite{cai2019pessimistic}, explores a new direction for \CEend.
}

\textit{ML-based query-driven \CE methods} try to learn a model to map each featurized query $Q$ to its cardinality $\Card(T, Q)$ directly. Some ML-enhanced methods improve the performance of \CE methods by using more complex models such as DNNs~\cite{kipf2018learned} or gradient boosted trees~\cite{dutt2019selectivity}.

\textit{ML-based data-driven \CE methods} are independent of the queries. They regard each tuple in $T$ as a point sampled according to the joint distribution $P_T(A) = P_T(A_1, A_2, \dots, A_n)$. 
Let $P_T(Q) = P_T(A_1 \in R_1, A_2 \in R_2, \cdots , A_n \in R_n)$ be the probability specified by the region of $Q$. Then, we have $\Card(T, Q) = P_T(Q) \cdot |T|$ so \CE problem could be reduced to model the probability density function (PDF) $P_T(A)$ of table $T$. 
A variety of ML-based models have been used in existing work to represent $P_T(A)$, the most representative of which includes deep auto-regression model~\cite{yang2019deep,yang2020neurocard,hasan2019multi} and probabilistic graphical models (PGMs) such as
Bayesian networks (BN)~\cite{tzoumas2011lightweight, getoor2001selectivity, wu2020bayescard}, SPN~\cite{hilprecht2019deepdb}, and FSPN~\cite{zhu2020flat}. 
\revise{
In addition, some methods proposed recently such as~\cite{wu2021unified} try to integrate both query and data information for \CEend.
}

\myskip
\noindent \underline{\textbf{\CE Benchmark:}}
Literature works have proposed some benchmark datasets and query workloads for \CE evaluation. We analyze their pros and cons as follows:

1) The synthetic benchmarks such as TPC-C~\cite{leis2015good}, TPC-H~\cite{tpch2021} and TPC-DS~\cite{tpcds2021} and Star Schema benchmarks (SSB)~\cite{o2009star} contain real-world data schemas and synthetic generated tuples. They are mainly used for evaluating query engines but not suitable for \CE because their data generator makes oversimplified assumptions on the joint PDF of attributes, such as uniform distribution and independence. 
However, real-world datasets are often highly skewed and correlated~\cite{leis2015good}, which are more difficult for \CE. 

2) IMDB dataset with its JOB workload~\cite{leis2015good} is a well-recognized benchmark, containing complex data and string ``LIKE'' queries. 
\revise{
To evaluate statistical \CE methods on n./c. queries only, most of the existing works~\cite{kipf2018learned,yang2019deep,yang2020neurocard,hilprecht2019deepdb,zhu2020flat} use the JOB-LIGHT query workload containing 70 selection queries with varied number of joining tables. However, these queries touch only 8 n./c. attributes within six tables of IMDB and the joins between these tables are only star joins centered at one table. 
Thus, this simplified IMDB dataset and its workload cannot comprehensively evaluate the performance of nowadays \CE algorithms on more complex data and varied join settings.} On the other hand, some works~\cite{yang2020neurocard, negi2021flow} generate queries on the IMDB dataset including ``LIKE'' queries which are not supported by most of recent statistical methods.

Apart from these well-established and general-purpose benchmarks, there also exist other benchmarks with specific purposes. For example, Wang~\cite{wang2020ready} presents a series of real-world datasets to analyze whether existing \CE algorithms are suitable to be deployed into real-world DBMS. However, it is only conducted on single-table datasets, which can not reflect the behavior of these models in more practical multi-table settings. 

\myskip
\noindent \underline{\textbf{Summary:}}
A surge of \CE algorithms built on top of statistical models has been proposed in the literature, especially in the last decade. However, existing \CE benchmarks are not sufficient to comprehensively evaluate their performance.


\section{Our New Benchmark}
\label{sec: bench}

In this section, we design a new benchmark with complex real-world data and diverse multi-table join query workload for evaluating \CE algorithms. To simulate practical scenarios, the benchmark should attain the following properties:

\indent \underline{\textit{1) Large scale}} with enough tables,  attributes, and  tuples in the full outer join;

\indent \underline{\textit{2) Complex distribution}} with skewed and correlated attributes whose joint distribution can not be modeled in a straightforward manner (e.g. independent assumption);

\indent \underline{\textit{3) Rich join schema}} containing joins of various number of tables and diverse join forms (e.g. star and chain);

\indent \underline{\textit{4) Diverse workload}} with queries covering a wide range of true cardinalities and different number of filtering and join predicates.

To this end, we establish our benchmark on a new real-world dataset with a hand-picked query workload. It overcomes the drawbacks of existing \CE benchmarks and fully fulfills the properties listed above. We describe the details on the data and workload settings in the following content. 

\begin{figure}
	\centering
	\includegraphics[width=6cm]{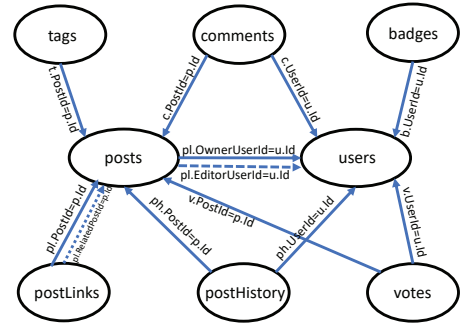}
	\vspace{-1em}
	\caption{Join relations between tables in STATS.}
	\label{fig: benchjoin}
	\vspace{-1em}
\end{figure}

\myskip
\noindent \underline{\textbf{Data Setting:}}
We adopt the real-world dataset STATS\footnote{\url{https://relational.fit.cvut.cz/dataset/Stats}}
in our benchmark. It is an anonymized dump of user-contributed content on the Stats Stack Exchange network. STATS consumes 658MB storage space with 8 tables and 71 \revise{n./c.} attributes on users, posts, comments, tags, and their relations. A comparison of the statistical information between STATS and IMDB \revise{(the simplified subset supporting JOB-LIGHT)} is shown in Table~\ref{tab: benchmarkstat-data}. We argue that STATS more suitable for \CE benchmark as follows:

\indent \underline{\textit{1) Larger scale:}} 
STATS has more data tables and a larger number of \revise{n./c. attributes} than \revise{the simplified} IMDB. Moreover, its full outer join size is four orders of magnitude larger.

\indent \underline{\textit{2) More complex data distribution:}} 
The distribution skewness of STATS and attribute correlation is more significant than \revise{the simplified} IMDB.
Moreover, STATS has $3 \times$ more attributes with a larger domain size, suggesting its PDF is much harder to model.

\indent \underline{\textit{3) Larger query space:}} 
\revise{
Each table in STATS has $1$ to $8$ n./c. attributes to be filtered while the simplified IMDB contains at most two in each table. 
}
Moreover, STATS's full outer join size is much larger than \revise{the simplified} IMDB.
These two aspects provide STATS a larger query space varying in cardinality and predicate numbers.

\indent \underline{\textit{4) Richer join settings:}} 
The join relations between all tables in STATS are shown in Figure~\ref{fig: benchjoin}.
\revise{The simplified} IMDB contains only star joins between primary key and foreign keys (i.e. 5 join relations). Whereas, STATS has richer and more diverse join types varying in the number of joined tables (from $2$ to $8$), join forms (chain, star, and mixture of these two), and join keys (PK-FK and FK-FK).

\begin{table}[t]
	\caption{Comparison of IMDB \revise{(simplified subset to fit JOB-LIGHT workload)} and STATS dataset.}
	\vspace{-1em}
	\scalebox{0.87}{
		\begin{tabular}{c|c|cc}
			\hline
			\rowcolor{mygrey}
			\textbf{Criteria} & \textbf{Item} & \textbf{IMDB} & \textbf{STATS} \\ \hline
			\multirow{4}{*}{Scale} &\# of tables & 6 & 8 \\
			&\# of \revise{n./c.} attributes & 8 & 23 \\
			&\# of \revise{n./c.} attributes per table & 1--2 & 1--8 \\
			&full outer join size & $2\cdot 10^{12}$ & $3\cdot10^{16}$ \\ \hline
			\multirow{3}{*}{Data}& total attribute domain size & 369, 563 & 578, 341 \\
			&average distribution skewness & 9.159 & 21.798 \\
			&average pairwise correlation & 0.149 & 0.221 \\ \hline
			\multirow{2}{*}{Schema} & join forms & star & star/chain/mixed\\
			& \# of join relations & 5 & 12\\ \hline
		\end{tabular}
		\label{tab: benchmarkstat-data}
	}
	\vspace{-2em}
\end{table}

\myskip
\noindent \underline{\textbf{Query Workload Setting:}}
We generate and then carefully handpick a query workload STATS-CEB on STATS to fulfill both practicality and diversity. The generation process is done in two phases.

In the first phase, we generate $70$ representative join templates based on the join schema in Figure~\ref{fig: benchjoin}, each of which specifies a distinct join pattern covering a set of tables. For these join templates, we do not consider: 
1) cyclic joins as most of the ML-based \CE algorithms~\cite{yang2020neurocard, hilprecht2019deepdb, zhu2020flat, wu2020bayescard} do not support them;
and 2) non-equal joins as they rarely occur in practice and many \CE algorithms process them in the same way as many-to-many joins.
We manually check and retain each join template if it has occurred in the log data of StackExchange
or has its real-world semantics.
To reduce redundancy, we also ensure that these join templates are not very similar (e.g. only differ in inner or outer join conditions).

In the second phrase after deriving these $70$ join templates, we generate $146$ queries with $1-4$ queries for each template as the testing query workload STATS-CEB. We make sure all the generated filter predicates reflect real-world semantics and diversify in multiple perspectives.
In comparison to JOB-LIGHT (illustrated in Table~\ref{tab: benchmarkstat-workload}), we find the following advantages of STATS-CEB:

\indent  \underline{\textit{1) More diverse queries:}} 
STATS-CEB contains twice queries as JOB-LIGHT with $3 \times$ more join templates covering a wider range of the number of joined tables.

\indent \underline{\textit{2) Richer join types:}}
Unlike JOB-LIGHT benchmark with only star joins, STATS-CEB contains queries with chain joins and complex mixed join forms. Moreover, JOB-LIGHT only contains queries with one-to-many PK-FK joins, whereas STATS-CEB includes queries with many-to-many FK-FK joins.

\indent \underline{\textit{3) More filter predicates:}}
STATS-CEB contains queries with up to 16 distinct filter predicates, which is $4 \times$ larger than JOB-LIGHT.

\indent \underline{\textit{4) Wider range of true cardinality:}} 
The cardinality range of STATS-CEB is an order of magnitude larger than JOB-LIGHT. The largest query in STATS-CEB has true cardinality of 20 billion, which is $3 \times$ larger than that of the JOB-LIGHT benchmark.

\myskip
\noindent \underline{\textbf{Summary:}}
\revise{
Our new benchmark with STATS dataset and STATS-CEB query workload are very comprehensive  with more complex data, more skewed distribution, more diverse queries and more complicated join settings. According to our evaluation results in the following Sections~5--7, this new \CE benchmark helps to find more insights over existing \CE algorithms that have not been discovered in previous works.
}

\begin{table}[t]
	\caption{Comparison of JOB-LIGHT and STATS-CEB benchmark query workload.}
	\vspace{-1em}
	\scalebox{0.9}{
		\begin{tabular}{c|cc}
			\hline
			\rowcolor{mygrey}
			\textbf{Item} & \textbf{JOB-LIGHT} & \textbf{STATS-CEB} \\ \hline
			\# of queries & 70 & 146 \\
			\# of joined tables & 2--5 & 2--8 \\
			\# of join templates & 23 & 70 \\
			\# of filtering \revise{n./c.} predicates & 1--4 & 1--16 \\
			join type & PK-FK & PK-FK/FK-FK\\
			true cardinality range & 9 --- $9\cdot10^9$ & 200 --- $2\cdot10^{10}$  \\ \hline
		\end{tabular}
	}
	\label{tab: benchmarkstat-workload}
	\vspace{-1em}
\end{table}


\section{Evaluation Plan}
\label{sec: plan}

We aim to evaluate how \CE algorithms behave in a real DBMS, including the end-to-end improvement on optimizing query plans and other practicality aspects, on our new benchmark. 
This section introduces the detailed evaluation plan. Section~\ref{sec: plan-algo} presents all baseline \CE algorithms chosen to be evaluated, Section~\ref{sec: plan-sys} describes our implementation method and system settings, and Section~\ref{sec: plan-met} lists the evaluation metrics of our interests.
 
 \subsection{\CE Algorithms}
 \label{sec: plan-algo}

\revise{
We identify and choose twelve representative \CE algorithms across the three classes (traditional, ML-based query-driven, and ML-based data-driven) reviewed in Section~\ref{sec: prelim}. The selection principles and details of algorithms in each class are elaborated as follows.
}


\noindent
\revise{
\underline{\textbf{Traditional \CE Algorithms:}}
In this class, we choose five algorithms along the technical directions:
1) for histogram-based methods, we evaluate \textsf{PostgreSQL} and \textsf{MultiHist}, which applies the one-dimensional and multi-dimensional histograms for \CEend;
2) for sampling-based methods, we evaluate the uniformly random sampling \textsf{UniSample} method and the more advanced \textsf{WJSample} method for join sampling;
and 3) for other methods, we evaluate \textsf{PessEst}, a recent proposed method that exhibit state-of-the-art performance in some aspects. 
The details are as follows:
}


\indent
\revise{1) \textsf{\underline{{PostgreSQL}}}~\cite{psql2020} refers to the histogram-based \CE method used in the well-known DBMS PostgreSQL. It assumes that all attributes are mutually independent and maintains a 1-D (cumulative) histogram to represent $P_{T}(A_i)$ for each attribute $A_i$. The probability $P_T(Q)$ can then be easily obtained by multiplying all $P_{T}(R_i)$ together. In addition, optimization strategies such as collecting the most common value and industrial implementation are used to enhance the performance.
}

\indent
\revise{
2) \textsf{\underline{{MultiHist}}}~\cite{poosala1997selectivity} identifies subsets of correlated attributes and model them as multi-dimensional histograms. We use the implementation provided in the repository~\cite{yang2019naru}. We do not compare with the variant methods DBHist~\cite{deshpande2001independence}, GenHist~\cite{gunopulos2000approximating,gunopulos2005selectivity} and 
VIHist~\cite{wang2003multi} over~\cite{poosala1997selectivity} since their improvements are not very significant and their open-sourced implementations are not provided.
}

\indent 3) \textsf{\underline{{\revise{UniSample}}}}~\cite{leis2017cardinality,zhao2018random} makes no assumption but randomly fetches records from $T$ on-the-fly according to $P_{T}(A)$ to estimate the probability $P_{T}(Q)$. It is also widely used in DBMS such as MySQL~\cite{mysql2020} and MariaDB~\cite{mdb2020}. We set the sampling size to $10^4$.

\indent
\revise{
4) \textsf{\underline{{\revise{WJSample}}}}~\cite{li2016wander} designs a random walk based method called wander join to sample tuples from multiple tables. It has been integrated into DBMSs and exhibits favorable performance in a recent study~\cite{park2020gcare}. The method~\cite{zhao2018random} then improves from biased to unbiased sampling. We do not compare with it to avoid redundancy.
}

\revise{
5) \textsf{\underline{{\revise{PessEst}}}}~\cite{cai2019pessimistic} leverages 
randomized hashing and data sketching to tighten the bound for multi-join queries. It is a new class of estimator as it never underestimates the cardinality. Meanwhile, it has been verified to perform well in real world DBMS.
}

\revise{
We do not compare with the other variants of traditional methods~\cite{bruno2001stholes, srivastava2006isomer, khachatryan2015improving, fuchs2007compressed, stillger2001leo, wu2018towards, heimel2015self,kiefer2017estimating, leis2017cardinality} as they do not exhibit significantly better performance or provide open-source implementation.}

\myskip

\noindent 
\revise{
\underline{\textbf{ML-Based \CE Algorithms:}}
In our evaluation, we choose four query-driven (\textsf{MSCN}, \textsf{LW-XGB}, \textsf{LW-NN} and \textsf{UAE-Q}) and four data-driven (\textsf{NeuroCard}, \textsf{BayesCard}, \textsf{DeepDB} and \textsf{FLAT}) \CE methods. They are representative as they apply different statistical models, and exhibit state-of-the-art performance using each model. Specifically, for query-driven methods, \textsf{MSCN}, \textsf{LW-XGB}/\textsf{LW-NN} and \textsf{UAE-Q} use deep neural networks, classic lightweight regression models and deep auto-regression models to learn the mapping functions, respectively. For the data-driven methods, they build the data distribution utilizing deep auto-regression models and three probabilistic graphical models: BN, SPN, and FSPN, respectively. They use different techniques to balance estimation accuracy, inference efficiency, and model size. Besides, we also evaluate \textsf{UAE}, an extension of \textsf{UAE-Q} using both query and data information.
The details are as follows:
}

\indent 6) \textsf{\underline{MSCN}}~\cite{kipf2018learned} is a deep learning method built upon the multi-set convolutional network model. The features of attributes in table $T$, join relations in query $Q$, and predicates of query $Q$ are firstly fed into three separate modules, where each is comprised of a two-layer neural network. Then, their outputs are averaged, concatenated, and fed into a final neural network for estimation. 
 
\indent 7) \textsf{\underline{LW-XGB}} and 8) \textsf{\underline{LW-NN}}~\cite{dutt2019selectivity}
 formulate the \CE mapping function as a regression problem and apply gradient boosted trees and neural networks for regression, respectively. Specifically, \textsf{LW-XGB} applies the XGBoost~\cite{chen2016xgboost} as it attains equal or better accuracy and estimation time than both LightGBM~\cite{ke2017lightgbm} and random forests~\cite{svetnik2003random} for a given model size.  As the original models only support single table queries, we extend them to support joins with an additional neural network to combine single-table information.

\indent 
\revise{
9) \textsf{\underline{UAE-Q}}~\cite{wu2021unified} applies the deep auto regression models to learn the mapping function. It proposes differentiable progressive sampling via the Gumbel-Sotfmax trick to enables deep auto-regression models to learn from queries.
}

For above query-driven \CE methods, we automatically generate $10^5$ queries as the training examples to train these models. 


10) \revise{\underline{\textsf{NeuroCard}}~\cite{yang2020neurocard}, the multi-table extension of \textsf{Naru}~\cite{yang2019deep}, is built upon a deep auto-regression model.}
It decomposes the joint PDF $P_{T}(A) = P_{T}(A_1) \cdot \prod_{i=2}^{k}  P_{T}(A_i|  A_1, A_2,  \dots, A_{i - 1})$ according to the chain rule and model each (conditional) PDF parametrically by a 4-layer DNN ($4 \times 128$ neuron units). All \revise{tables} can be learned together using a \revise{single} masked auto-encoder~\cite{made}. Meanwhile, a progressive sampling technique~\cite{liang2020variable} is provided to sample points from the region of query $Q$ to estimate its probability. We set the sampling size to $8, 000$. 
We omit a very similar method in~\cite{hasan2019multi} as it has \revise{slightly worse} performance than \textsf{NeuroCard}. 
\revise{Worth noticing that the original \textsf{NeuroCard} method is only designed for datasets with tree-structured join schema. On our STATS benchmark with cyclic join schema, we partition the schema into multiple tree-structured join schemas and build one \textsf{NeuroCard} model for each schema. To avoid ambiguity, we denote this extension method as \textsf{NeuroCard}$^E$ in the following content.}


11) \textsf{\underline{BayesCard}}~\cite{wu2020bayescard} is fundamentally based on BN, which models the dependence relations among all attributes as a directed acyclic graph. Each attribute $A_i$ is assumed to be 
conditionally independent of the remaining ones given its parent attributes $A_{\text{pa}(i)}$ so the joint PDF $\Pr_{T}(A) = \prod_{i = 1}^{k} \Pr_{T}(A_i | A_{\text{pa}(i)})$. 
\textsf{BayesCard} revitalizes \textsf{BN} using probabilistic programming to improve its inference and model construction speed (i.e., learning the dependence graph and the corresponding probability parameters). Moreover, it adopts the advanced ML techniques to process the multi-table join queries, which significantly increases its estimation accuracy over previous BN-based \CE methods~\cite{getoor2001selectivity, tzoumas2011lightweight, dasfaa2019}, which will not be evaluated in this paper. We use the Chow-Liu Tree~\cite{chow1968approximating} based method to build the structure of \textsf{BayesCard} and apply the complied variable elimination algorithm for inference.

\indent 12) \textsf{\underline{DeepDB}}~\cite{hilprecht2019deepdb}, based on sum-product networks (SPN)~\cite{poon2011sum,desana2020sum}, approximates $P_T(A)$ by recursively decomposing it into local and simpler PDFs. Specifically, the tree-structured SPN contains sum node to split $P_T(A)$ to multiple $P_{T'}(A)$ on tuple subset $T' \subseteq T$, product node to decompose $P_{T}(A)$ to $\prod_{S}P_{T}(S)$ for independent set of attributes $S$ and leaf node if $P_{T}(A)$ is a univariate PDF. The SPN structure can be learned by splitting table $T$ in a top-down manner. Meanwhile, the probability of $\Pr_{T}(Q)$ can be obtained in a bottom-up manner with time cost linear in the SPN's node size. 

\indent 13) \textsf{\underline{FLAT}}~\cite{zhu2020flat}, based on factorize-split-sum-product networks (FSPN)~\cite{wu2020fspn}, improves over SPN by
adaptively decomposing $P_T(A)$ according to the attribute dependence level. It adds the factorize node to split $P_T$ as $P_T(W) \cdot P_T(H | W)$ where $H$ and $W$ are highly and weakly correlated attributes in $T$. $P_T(W)$ is modeled in the same way as SPN. $P_T(H | W)$ is decomposed into small PDFs by the split nodes until $H$ is locally independent of $W$. Then, the multi-leaf node is used to model the multivariate PDF $P_T(H)$ directly. Similar to SPN, the FSPN structure and query probability can be recursively obtained in a top-down and bottom-up fashion, respectively. 

For both \textsf{DeepDB} and \textsf{FLAT}, we set the RDC thresholds to $0.3$ and $0.7$ for filtering independent and highly correlated attributes, respectively. Meanwhile, we do not split a node when it contains less than $1\%$ of the input data.


\indent 
\revise{
14) \textsf{\underline{UAE}}~\cite{wu2021unified} extends the \textsf{UAE} method by unifiying both query and data information using the auto-regression model. It is a representative work aiming at closing the gap between data-driven and query-driven \CE methods.
}

\myskip

\noindent 
\underline{\textbf{Remarks:}}
For each \CE algorithm, we adopt the publicly available implementation~\cite{hilp2019deepdb, yang2019naru, yang2020sb} if the authors provide it and otherwise implement it by ourselves. For other hyper-parameters, if they are known to be a trade-off of some metrics, we choose the default values recommended in the original paper. Otherwise, we run a grid search to explore the combination of their values that largely improves the end-to-end performance on a validation set of queries. Notice that, each of our evaluated \CE algorithms is an independent and universal tool that can be easily integrated into common DBMS. There have also been proposed some \CE modules~\cite{sun2019end, wu2021unified} that are optimized together with other components in a query optimizer in an end-to-end manner. We do not compare with them as they do not fit our evaluation framework.

\subsection{Implementation and System Settings}
\label{sec: plan-sys}

To make our evaluation more realistic and convincing, we integrate each \CE algorithm into the query optimizer of PostgreSQL~\cite{psql2020}, a well-recognized open-source DBMS. Then, the quality of each \CE method can be directly reflected by the end-to-end query runtime with their injected cardinality estimation.

Before introducing the details of our integration strategy, we introduce an important concept called \textit{sub-plan query}. For each SQL query $Q$, each \textit{sub-plan query} is a query touching only a subset of tables in $Q$.
The set of all these queries is called \textit{sub-plan query space}. For the example query $A \bowtie B \bowtie C$, its sub-plan query space contains queries on $A \bowtie B$, $A \bowtie C$, $B \bowtie C$, $A$, $B$, and $C$ with the corresponding filtering predicates. The built-in planner in DBMS will generate the sub-plan query space, estimate their cardinalities, and determine the optimal execution plan. For example, the sub-plan queries $A$, $B$, and $C$ only touch a single table, their \CE results may affect the selection of table-scan methods, i.e. index-scan or seq-scan. The sub-plan queries $A \bowtie B$, $A \bowtie C$, and $B \bowtie C$ touch two tables. Their cardinalities may affect the join order, i.e. joining $A \bowtie B$ with $C$ or $A \bowtie C$ with $B$, and the join method, i.e. nested-loop-join, merge-join, or hash-join. Therefore, the effects of a \CE method on the final query execution plan are entirely decided by its estimation results over the sub-plan query space.

To this end, in our implementation, 
we overwrite the function ``\textsf{calc\_joinrel\_size\_estimate}'' in the planner of PostgreSQL to derive the sub-plan query space for each query in the workload. 
Specifically, every time the planner needs a cardinality estimation of a sub-plan query, the modified function ``\textsf{calc\_joinrel\_size\_estimate}'' will immediately capture it.
Then, we call each \CE method to estimate the cardinalities of the sub-plan queries and inject the estimations back into PostgreSQL. Afterward, we run the compiler of PostgreSQL on $Q$ to generate the plan. It will directly read the injected cardinalities produced by each method. Finally, we execute the query with the generated plan. In this way, we can support any \CE method without a large modification on the source code of PostgreSQL. We can report the total time (except the sub-plan space generation time) as the end-to-end time cost of running a SQL query using any \CE method.

For the environment, we run all of our experiments in two different Linux Servers. The first one 
with 32 Intel(R) Xeon(R) Platinum 8163 CPUs @ 2.50GHz, one Tesla V100 SXM2 GPU and 64 GB available memory is used for model training. The other one with 64 Intel(R) Xeon(R) E5-2682 CPUs @ 2.50GHz is used for the end-to-end evaluation on PostgreSQL.

 \subsection{Evaluation Metrics}
 \label{sec: plan-met}
 
Our evaluation mainly focuses on \emph{quantitative} metrics that directly reflect the performance of \CE algorithms from different aspects. We list them as follows:

\indent 1) \text{\underline{End-to-end time}} of the query workload, including both the query plan generation time and physical plan execution time. It serves as the ``gold-standard'' for \CE algorithm, since improving the end-to-end time is the ultimate goal for optimizing \CE in query optimizers. 
\revise{
We report the end-to-end time of \textsf{TrueCard}, which injects the true cardinalities of all sub-plan queries into PostgreSQL. Ideally if the cost model is very accurate, \textsf{TrueCard} can obtain the optimal plan with shortest time. For a fixed PostgreSQL cost model, we find \textsf{TrueCard} can obtain the optimal query plan for most of the time. Thus, this could serve as a good baseline.
}

\indent 2) \text{\underline{Inference latency}} reflects the time cost for \CE, which directly relates to the query plan generation time. It is crucial as \CE needs to be done numerous times in optimizing the plan of each query. Specifically, an accurate \CE method may be very time-costly in inference. Despite the fast execution time of the plans generated by this method, the end-to-end query performance can be poor because of its long plan generation time.

\indent 3) \text{\underline{Space cost}} refers to the \CE model size. A lightweight model is also desired as it is convenient to transfer and deploy.

\indent 4) \text{\underline{Training cost}} refers to the models' offline training time. 

\indent 5) \text{\underline{Updating speed}} reflects the time cost for \CE models update to fit the data changes. For real-world settings, this metric plays an important role as its underlying data always updates with tuples insertions and deletions.

Besides these metrics, \cite{wang2020ready} proposed some \emph{qualitative metrics} related to the stability, usage, and deployment of \CE algorithms and made a comprehensive analysis. Thus, we do not consider them in this paper.
In the following, we first evaluate the overall end-to-end performance of all methods in Section~\ref{sec: static}. Then, we analyze the other practicality aspects in Section~\ref{sec: dynamic}. At last, we point out the drawbacks of existing evaluation metric and propose a new metric as its potential substitution in Section~\ref{sec:analysis}.

	\section{How Good Are The \CE Methods?}

\label{sec: static}
In this section, we first thoroughly investigate the true effectiveness of the aforementioned \CE methods in improving query plan quality. 
Our evaluation focuses on a static environment where data in the system has read-only access. This setting is ubiquitous and critical for commercial DBMS, especially in OLAP workloads of data warehouses\cite{nam2020sprinter,garcia1982read,pedreira2018rethinking,zhao2020efficient}. 
We organize the experimental results as follows: Section~\ref{sect5.1} reports the overall evaluation results, Section~\ref{sect5.2} provides detailed analysis of the method's performance on various query types and an in-depth case study on the performance of some representative methods. 

\subsection{Overall End-to-End Performance }
\label{sect5.1}

\begin{table*}
	\centering
	\caption{Overall performance of \CE algorithms.}
	\vspace{-0.5em}
	\resizebox{0.95\textwidth}{!}{
		\begin{tabular}{cc|ccc|ccc}
			\toprule
			\rowcolor{mygrey}
			&  & \multicolumn{6}{c}{\textbf{Workload}} \\
			\rowcolor{mygrey}
			\multirow{-2}{*}[-2ex]{\textbf{Category}} & 
			\multirow{-2}{*}[-2ex]{\textbf{Method}} & \multicolumn{3}{c}{\textsf{\textbf{JOB-LIGHT}}} & \multicolumn{3}{c}{\textsf{\textbf{STATS-CEB}}}\\ 
			\cmidrule[0.05em](lr){3-5} \cmidrule[0.05em](l){6-8}
			\rowcolor{mygrey}
			& & \textbf{End-to-End Time} & \textbf{Exec. + Plan Time} & \textbf{Improvement}  & \textbf{End-to-End Time} & \textbf{Exec. + Plan Time} & \textbf{Improvement} \\
			\midrule
			\multirow{2}{*}{Baseline} &
			\textsf{PostgreSQL} & 3.67h & 3.67h + 3s  & $0.0\%$ & 11.34h & 11.34h + 25s & $0.0\%$ \\
			& \textsf{\textbf{TrueCard}} & \textbf{3.15h} & \textbf{3.15h + 3s} & $\mathbf{14.2\%}$ &  \textbf{5.69h} & \textbf{5.69h + 25s} & $\mathbf{49.8\%}$ \\ \hline
			\multirow{4}{*}{Traditional} & \revise{\textsf{MultiHist}} & \revise{3.92h} & \revise{3.92h + 30s} & $-6.8\%$  & \revise{14.55h} & \revise{14.53h + 79s} & \revise{$-28.3\%$}  \\
			& \revise{\textsf{UniSample}} & 4.87h & 4.84h + 96s & $-32.6\%$& $>25h$ & $--$ & $--$\\
			& \revise{\textsf{WJSample}} & \revise{4.15h} & \revise{4.15h + 23s} & \revise{$-13.1\%$}& \revise{19.86h} & \revise{19.85h + 45s} & \revise{$-75.0\%$}\\
			& \revise{\textsf{PessEst}} & \revise{3.38h} & \revise{3.38h + 11s} & \revise{$7.9\%$}& \revise{6.10h} & \revise{6.10h + 43s} & \revise{$46.2\%$}\\
			\hline
			\multirow{4}{*}{Query-driven} & \textsf{MSCN} & \revise{3.50h} & \revise{3.50h + 12s} & \revise{$4.6\%$}  & \revise{$8.13h$} & \revise{8.11h + 46s}
			& \revise{$28.3\%$} \\
			& \textsf{LW-XGB} & 4.31h & 4.31h + 8s & $-17.4\%$  & $>25h$ & $--$ & $--$\\ 
			& \textsf{LW-NN} & 3.63h & 3.63h + 9s & $1.1\%$  & 11.33h & 11.33h + 34s & $0.0\%$ \\ 
			& \revise{\textsf{UAE-Q}} & \revise{3.65h}  & \revise{3.55h+356s} & \revise{$-1.9\%$} & \revise{11.21h} & \revise{11.03h+645s} & \revise{$1.1\%$} \\\hline
			
			\multirow{4}{*}{Data-driven}
			& \textsf{NeuroCard}$^E$ & 3.41h & 3.29h + 423s & $6.8\%$ & 12.05h & 11.85h + 709s & $-6.2\%$ \\
			& \textsf{BayesCard} & 3.18h & 3.18h + 10s & $13.3\%$  & 7.16h & 7.15h + 35s & $36.9\%$ \\
			& \textsf{DeepDB} & 3.29h & 3.28h + 33s &  $10.3\%$  & 6.51h & 6.46h + 168s &  $42.6\%$ \\
			& \textsf{FLAT} & 3.21h & 3.21h + 15s &  $12.9\%$  & 5.92h & 5.80h + 437s &  $47.8\%$ \\ \hline
			Query + Data & \revise{\textsf{UAE}} & \revise{3.71h} & \revise{3.60h + 412s} & \revise{$-2.7\%$} & \revise{11.65h}  & \revise{11.46h + 710s} & \revise{$-0.02\%$} \\ 
			\bottomrule
		\end{tabular}
	}
	\label{tab:overall}
\end{table*}

We evaluate the end-to-end performance (query execution time plus planning time) on both JOB-LIGHT and STATS-CEB benchmarks for all \CE methods including two baselines \textsf{PostgreSQL} and \textsf{TrueCard} shown in Table~\ref{tab:overall}.
We also report their relative improvement over the \textsf{PostgreSQL} baseline as an indicator of their end-to-end performance. In the following, we will first summarize several overall observations (O) regarding Table~\ref{tab:overall}, and then provide detailed analysis w.r.t. each of these \CE methods.

\smallskip
\noindent \revise{\textbf{O1: Most of the ML-based data-driven \CE methods can achieve remarkable performance, whereas most of the traditional and  ML-based query-driven \CE methods do not have much improvement over \textsf{PostgreSQL}.}}
The astonishing performance of these ML-based data-driven \CE methods \break (\textsf{BayesCard}, \textsf{DeepDB}, and \textsf{FLAT}) come from their accurate characterization of data distributions and \revise{reasonable independence assumption over joined tables.} \revise{Traditional histogram and sampling based methods (\textsf{MultiHist}, \textsf{UniSample}, and \textsf{WJSample}) have worse performance than \textsf{PostgreSQL} whereas the new traditional approach (\textsf{PessEst}) is significantly better. The query-driven \CE methods' performance is not stable. They
rely on a large amount of executed queries as training data and the testing query workload should follow the same distribution as the training workload to produce an accurate estimation~\cite{hilprecht2019deepdb}.}



\smallskip
\noindent \revise{\textbf{O2: The differences among the \CE methods' improvements over \textsf{PostgreSQL} are much more drastic on datasets with more complicated data distributions and join schemas.}}
\revise{We observe that the execution time for \CE method that can outperform \textsf{PostgreSQL} (\textsf{PessEst}, \textsf{NeuroCard}$^E$, \textsf{BayesCard}, \textsf{DeepDB}, and \textsf{FLAT}) on JOB-LIGHT are all roughly $3.2h$, which is very close to the minimal execution time of \textsf{TrueCard}($3.15h$).}
As explained in Section~\ref{sec: bench}, the data distributions in \revise{the simplified IMDB dataset} and the JOB-LIGHT queries are relatively simple. Specifically, the table \textsf{title} in the IMDB dataset plays a central role in the join schema that other tables are all joined with its primary key, \revise{so the joint distribution could be easily learned}. However, their performance differences on STATS are very drastic
because the STATS dataset is much more challenging with high attribute correlations and various join types. 
Therefore, the STATS-CEB benchmark \revise{can help} expose the advantages and drawbacks of these methods.


\smallskip
\noindent \underline{\textbf{Analysis of Traditional \CE Methods:}}
\revise{Histogram and sampling based methods perform significantly worse than \textsf{PostgreSQL} on both benchmarks because of their inaccurate estimation. 
\textsf{MultiHist} and \textsf{UniSample} use the join uniformity assumption to estimate join queries, whose estimation error grows rapidly for queries joining more tables.
\textsf{WJSample} makes a random walk based sample across the join of multiple tables. However, as the cardinality increases with the number of joined tables, the relatively small sample size can not effectively capture the data distribution, leading to large estimation error.
These queries joining larger number of tables are generally more important in determining a good execution join order~\cite{leis2015good}.
Therefore, these methods tend to yield poor join orders and long-running query plans. 
The \textsf{PostgreSQL} produces more accurate estimations because of its high-quality implementation and fine-grained optimizations on join queries.
The new traditional method \textsf{PessEst} has a significant improvement over the \textsf{PostgreSQL} because it can compute the upper bound on estimated cardinalities to avoid expensive physical join plans.
}

\smallskip
\noindent \underline{\textbf{Analysis of ML-based Query-driven \CE Methods:}}
Overall the query-driven methods have comparable performance to the \textsf{PostgreSQL} baseline. Specifically, \textsf{MSCN} can slightly outperform the \textsf{PostgreSQL} (4.6\% faster runtime on JOB-LIGHT and 19.7\% faster on STATS-CEB), \textsf{LW-XGB} has much slower query runtime, and \textsf{LW-NN} has comparable performance. 
The unsatisfactory performance of these methods could be due to the following reasons. 

$\bullet$ These methods are essentially trying to fit the probability distributions of all possible joins in the schema, which has super-exponential complexity. Specifically, there can exist an exponential number possible joins in a schema, and for each joining table, the complexity of its distribution is exponential w.r.t. its attribute number and domain size~\cite{wu2020bayescard}.
Thus, these models themselves are not complex enough to fully understand all these distributions. 

$\bullet$ Similarly, these methods would require an enormous amount of executed queries as training data to accurately characterize these complex distributions. In our experiment, our computing resources can only afford to generate $10^5$ queries (executing 146 queries in STATS-CEB takes 10 hours), which may not be enough for this task. Besides, it is unreasonable to assume that a \CE method can have access to this amount of executed queries in reality.

$\bullet$ The well-known workload shift issue states that query-driven methods trained one query workload will not likely produce an accurate prediction on a different workload~\cite{hilprecht2019deepdb}. In our experiment, the training query workload is automatically generated whereas the JOB-LIGHT and STATS-CEB testing query workload is hand-picked. Therefore, the training and testing workload of these methods have different distributions. 

\smallskip
\noindent \underline{\textbf{Analysis of ML-based Data-driven Methods:}}
Data-driven ML methods (\textsf{BayesCard}, \revise{\textsf{NeuroCard}$^E$}, \textsf{DeepDB}, and \textsf{FLAT}), do consistently outperform \textsf{PostgreSQL} by $7-13\%$ on JOB-LIGHT. Except for \revise{\textsf{NeuroCard}$^E$}, the other three improve the \textsf{PostgreSQL} by $37-48\%$ \revise{on STATS-CEB}.
\revise{Their performance indicates that data-driven methods could serve as a practical counterpart of the PostgreSQL \CE component.} Through detailed analysis of \revise{\textsf{NeuroCard}$^E$} method, we derive the following observation:

\smallskip
\noindent \revise{\textbf{O3: Learning one model on the (sample of) full outer join of all tables in a DB may lead to poor scalability. We conjecture that an effective \CE method should make appropriate independent assumptions for large datasets.}}
The advantages of \revise{\textsf{NeuroCard}$^E$} over \textsf{PostgreSQL} disappear when shifting from JOB-LIGHT to STATS-CEB benchmark for the following reasons. First, the STATS dataset contains significantly more attributes with larger domain size, which \revise{can be} detrimental to \revise{\textsf{NeuroCard}$^E$}'s underlying deep auto-regressive models~\cite{wang2020ready, wu2020bayescard}. Second, the full outer join size of STATS is significantly larger than \revise{the simplified} IMDB, making the sampling procedure of \revise{\textsf{NeuroCard}$^E$} less effective. Specifically, the full outer join size can get up to $3\times 10^{16}$ and an affordable training data sample size would be no larger than $3\times 10^8$. Therefore, the \revise{\textsf{NeuroCard}$^E$} model trained on this sample only contains $1\times 10^{-8}$ of the information as the original dataset.
\revise{Third, the join keys in STATS dataset have very skewed distribution. E.g. there exist key values of a table that can match with zero or one as well as hundreds of tuples in another table. This complicated distribution of join keys makes \textsf{NeuroCard}$^E$'s full outer join sample less effective.}
\revise{Therefore, \textsf{NeuroCard}$^E$} can hardly capture the correct data distributions especially for join tables with small cardinalities. Specifically, we find that for queries on the joins of a small set of tables, \revise{\textsf{NeuroCard}$^E$}'s prediction deviates significantly from the true cardinality because its training sample does not contain much not-null tuples for this particular set of join tables. 

\begin{table}[t]
	\caption{End-to-end time improvement ratio of \CE algorithms on queries with different number of join tables.}
	\vspace{-1em}
	\scalebox{0.69}{
		\begin{tabular}{cc|cccccc}
			\hline
			\rowcolor{mygrey}
			\bf \# tables  & \bf \# queries & \bf \revise{\textsf{PessEst}} & \bf \textsf{MSCN} & \bf \textsf{BayesCard} & \bf \textsf{DeepDB} & \bf \textsf{FLAT} & \textbf{TrueCard}\\ \hline
			{$2-3$} & 38 & \revise{$2.62\%$} & \revise{$2.04\%$} & $2.07\%$ & $1.98\%$ & $2.48\%$  & $\mathbf{3.66\%}$  \\
			{$4$} & 50 & \revise{$53.1\%$} & \revise{$-12.3\%$} & $55.8\%$ & $48.0\%$ & $55.7\%$ & $\mathbf{55.9\%}$\\
			{$5$} & 28 & \revise{$31.7\%$} & \revise{$29.8\%$} & $36.55\%$  & $32.90\%$ & $35.4\%$ & $\mathbf{37.0\%}$ \\ 
			{$6-8$} & 34 & \revise{$29.6\%$} & \revise{$-4.06\%$} & $2.51\%$ & $26.3\%$ & $32.0\%$ & $\mathbf{34.6\%}$ \\
			\hline
	\end{tabular}}
	\label{tab: join_number}
	\vspace{-2em}
\end{table}

All other three data-driven \CE methods can significantly outperform the \textsf{PostgreSQL} baseline because their models are not constructed on the full outer join of all tables. 
Specifically, they all use the ``divide and conquer'' idea to divide the large join schema into several smaller subsets with each representing a join of multiple tables. 
In this way, they can capture the rich correlation within each subset of tables;
simultaneously, they avoid constructing the full outer join of all tables by assuming some independence between tables with low correlations.
Then, \textsf{BayesCard}, \textsf{DeepDB}, and \textsf{FLAT} build a model (BN, SPN, and FSPN respectively) to represent the distribution of the corresponding small part. This approach solves the drawback of \revise{\textsf{NeuroCard}$^E$}, yields relatively accurate estimation, and produces effective query execution plans. 
Among them, \textsf{FLAT} achieves the best performance (47.8\% improvement), which is \revise{very close to the improvement 49.8\% for \textsf{TrueCard}}. It can outperform \textsf{DeepDB} mostly because the STATS dataset is highly correlated, so the FSPN in \textsf{FLAT} has a more accurate representation of the data distribution than the SPN in \textsf{DeepDB}. On the other hand, \textsf{BayesCard} has an even more accurate representation of data distribution and yields the best end-to-end time for most queries in STATS-CEB. It does not outperform \textsf{FLAT} most because of one extremely long-run query, which we will study in detail in Section~\ref{sect5.2}.

\subsection{Analysis of Different Query Settings}
\label{sect5.2}

In this section, we further examine to what extent the \CE methods improve over \textsf{PostgreSQL} on various query types, i.e. different number of join tables ($\# tables$) and different intervals of true cardinalities. 
Since JOB-LIGHT workload does not contain queries with very diverse types and the ML-based data-driven methods do not show significant difference on these queries, we only investigate queries on STATS-CEB.
Worth noticing that we only examine the methods with clear improvements over \textsf{PostgreSQL} on STATS-CEB: \textsf{MSCN}, \textsf{BayesCard}, \textsf{DeepDB}, and \textsf{FLAT}.

\smallskip
\noindent \textbf{\underline{Number of Join Tables:}}
Table~\ref{tab: join_number} shows performance improvement of different ML-based methods over the \textsf{PostgreSQL} baseline and we derive the following observation:

\noindent\textbf{\revise{O4: The improvement gaps between these methods and the performance of \textsf{TrueCard} increase with the number of join tables.}}
Specifically, the \textsf{BayesCard} achieves near-optimal improvement for queries joining no more than 5 tables, but it barely has much improvement for queries joining 6 tables and more.
This observation suggests that the estimation qualities of these SOTA methods decline for queries joining more tables. 
In fact, the fanout join estimation approach adopted by all these methods sacrifices accuracy for efficiency by assuming some tables are independent of others. This estimation error may accumulate for queries joining a large number of tables, leading to sub-optimal query plans.

\begin{figure*}[htbp]
	\centering
	\includegraphics[width=16cm]{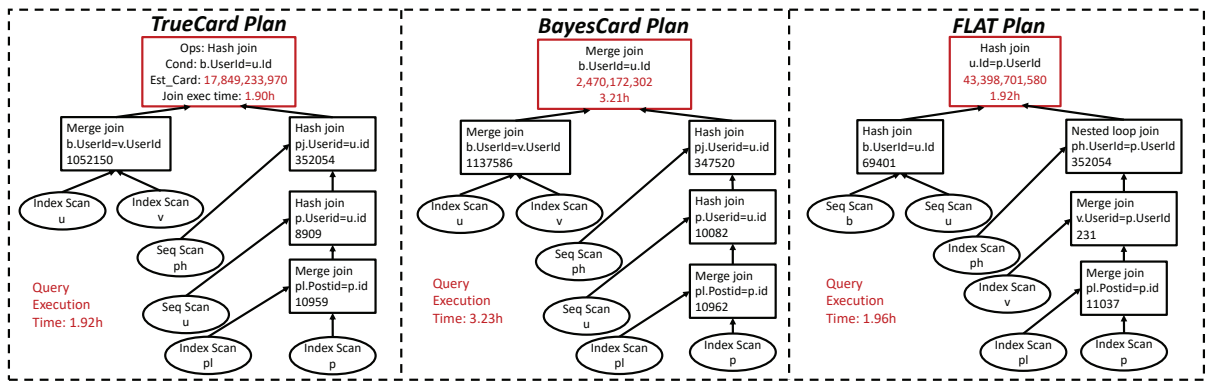}
	\vspace{-1em}
	\caption{\revise{Case study of STATS Q57.}}
	\label{fig:q57plans}
	\vspace{-1em}
\end{figure*}

\smallskip
\noindent \textbf{\underline{Size of Cardinality:}}
\revise{We choose Q57 (in Figure~\ref{fig:q57plans}) of STATS-CEB as a representative query to study the effect of estimation accuracy w.r.t. different cardinalities and investigate when a \CE method could go wrong. }
The execution time of Q57 for \textsf{TrueCard} and \textsf{FLAT} is $1.90h$ and $1.92h$, while the time for \textsf{BayesCard} is $3.23h$. We derive two important observations from this query, which are verified to be generalizable to other queries in JOB-LIGHT and STATS-CEB.

\smallskip
\noindent \revise{ \textbf{O5: Accurate estimation of (sub-plan) queries with large cardinalities is sometimes more important than the small ones.}}
When choosing the join method in the root node of execution plans for Q57, \textsf{BayesCard} underestimates the final join size and chooses the ``merge join'' physical operation. Alternatively, \textsf{FLAT} produces a more accurate estimation for the final join size and chooses the ``hash join'' operation, which is twice as faster as the ``merge join''. Since the final join operation takes up $99\%$ of the total execution time, \textsf{FLAT} significantly outperforms \textsf{BayesCard} on this query.

Generally, the (sub-plan) query with larger true cardinality requires a longer time to execute. It is very common that the join size of two intermediate tables is much larger than both of them. Therefore, some sub-plans can take a significantly longer time to execute than other sub-plans. A bad estimation on these large sub-plan queries can have a detrimental result on the overall runtime, whereas a series of good estimations on small sub-plan queries will not influence the runtime as much. Therefore, the estimation accuracy of sub-plan queries with very large true cardinalities dominate the overall quality of the query plan.

\smallskip
\noindent \textbf{O6: Choosing the correct physical operations sometimes is more important than selecting the optimal join order.} 
As shown in Figure~\ref{fig:q57plans} \textsf{BayesCard} can generate the optimal join order of Q57 because of its near-perfect estimation of all sub-plan queries except for the one at the root node.
The join order selected by \textsf{FLAT} is very different from the optimal one. Surprisingly, \textsf{FLAT}'s plan is roughly twice faster to execute than \textsf{BayesCard}'s plan due to the dominant large sub-plan query at the root node. For Q57, a sub-optimal query plan is only $1\%$ slower to execute but only one sub-optimal physical operation is $68\%$ slower. 

These aforementioned two observations also hold for other queries in these two benchmarks, so we conjecture that they might be generalizable to all queries. 

\section{What Other Aspects of \CE Methods Matter?}
\label{sec: dynamic}

In addition to \CE's improvement in execution time, we discuss model practicality aspects in this section: inference latency (in Section~\ref{sect6.1}), model size and training time (in Section~\ref{sect6.2}), and model update speed and accuracy (in Section~\ref{sect6.3}). 
\revise{We only compare the recently proposed \CE methods, which have been proved to significantly improve the \textsf{PostgreSQL} baseline, namely \textsf{PessEst}, \textsf{MSCN}, \textsf{NeuroCard}$^E$, \textsf{BayesCard}, \textsf{DeepDB}, and \textsf{FLAT}.}

\revise{
\begin{table}
	\caption{\revise{OLTP/OLAP Performance on STATS-CEB.}}
	\vspace{-1.2em}
	\label{tab:TPAP}
	\scalebox{0.71}{
		\begin{tabular}{ccc|cc}
			\hline
			\rowcolor{mygrey}
			\revise{\bf Methods} & \revise{\bf \textsf{TP Exec. Time}} & \revise{\bf \textsf{TP Plan Time}} & \revise{\bf \textsf{AP Exec. Time}} & \revise{\bf \textsf{AP Plan Time}} \\ \hline
			 \textsf{PostgreSQL}& 44.7s & 4.8s (9.7\%) & 11.32h & 20.3s ($0.05\%$)   \\ 
			 \textsf{TrueCard} & 8.2s & 4.8s (36.9\%) & 5.68h & 20.3s ($0.1\%$) \\ 
			 \textsf{PessEst} & 19.3s & 8.4s (30.3\%) & 6.09h & 35.4s ($0.16\%$) \\ 
			 \textsf{MSCN} & 15.7s & 8.2s (34.3\%) & 8.11h & 38.0s ($0.13\%$) \\ 
			 \textsf{NeuroCard$^E$} & 26.3s & 73s (73.5\%) & 11.84h & 350s ($0.81\%$) \\ 
			 \textsf{BayesCard} & 10.7s & 7.3s (40.6\%) & 7.15h & 27.4s ($0.11\%$) \\ 
			 \textsf{DeepDB}    & 11.5s & 33.6s (74.5\%) & 6.46h & 135s ($0.58\%$) \\ 
			 \textsf{FLAT}  & 14.3s & 41.5s (74.4\%) & 5.80h & 396s ($1.86\%$) \\ \hline
  		\end{tabular}
	}
	\vspace{-2em}
\end{table}}

\subsection{Inference Latency}
\label{sect6.1}

The end-to-end query time is comprised of query execution and planning time, the latter of which is determined by the \CE method's inference latency. Commercial DBMS normally has a negligible planning time due to their simplified cardinality estimator and engineering effort to accelerate the inference speed. However, the inference latency of some ML-based data-driven methods can approach one second per sub-plan query, which slows down the end-to-end query execution time. 
\revise{To further illustrate the importance of inference latency, we divide the STATS-CEB queries into OLTP and OLAP two workloads based on the query execution time. We report the results in Table~\ref{tab:TPAP} and derive the following observation.}

\smallskip
\noindent \revise{\textbf{O7: Inference latency can have a significant impact on the OLTP workload but a trivial impact on the OLAP workload.}
On OLTP workload of STATS-CEB, we observe that the planning time composes a large proportion of total end-to-end time. Specifically, some ML-based methods' (\textsf{NeuroCard}$^E$, DeepDB, and FLAT) inference speeds are relatively slow. Although their execution time on OLTP workload is faster than \textsf{PostgreSQL}, they have worse end-to-end performance because of the long planning time. 
For OLAP workload of STATS-CEB, the \CE methods' planning time is much shorter than their execution time because OLAP workload contains extremely long-run queries. In this case, the quality of the generated query plans overshadows the slow inference latency.
Therefore, we believe that \CE methods targeting different workloads should fulfill different objectives. For OLTP workload, a desirable method should have fast inference speed, whereas the methods targeting OLAP workload can have high inference latency as long as they can produce high-quality query plans.}



\begin{figure}
	\centering
	\includegraphics[width=\columnwidth]{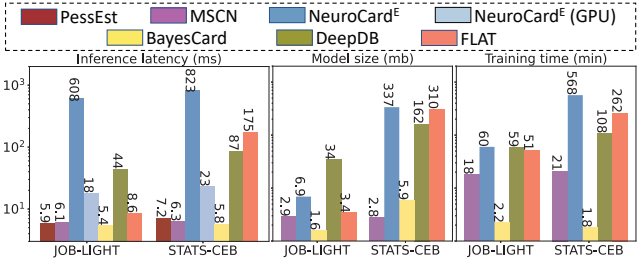}
	\vspace{-2.5em}
	\caption{\revise{Practicality aspects of \CE algorithms.}}
	\vspace{-2.5em}
	\label{fig: practicality}
\end{figure}

Figure~\ref{fig: practicality} reports the average inference latencies of all sub-queries in the workload for each method. 
Their inference speed can be ranked as \textsf{BayesCard} $>$ \revise{\textsf{NeuroCard}$^E$(GPU)} $>$ \textsf{FLAT}/\textsf{DeepDB} $>>$ \revise{\textsf{NeuroCard}$^E$}. 
The newly proposed inference algorithms on BN provide \textsf{BayesCard} with a very fast and stable inference speed on both benchmarks. 
However, the inference speeds of \textsf{FLAT} and \textsf{DeepDB} are not as stable because they tend to build much larger models with more computation circuits for the more complicated database STATS.
The inference process of \textsf{NeuroCard} requires a large number of progressive samples and its underlying DNN is computationally demanding on CPUs. Therefore, we observe that the inference speed is greatly improved by running it on GPUs. 

\subsection{Model Deployment}
\label{sect6.2}
\revise{Figure~\ref{fig: practicality} reports the model size and training time of all aforementioned methods. We do not report \textsf{PessEst} because it is a model-free method that does not need training. 
Based on the results of STATS-CEB query, we derive the following observation.}

\noindent \textbf{O8: BN-based \CE approach is very friendly for system deployment.}
\revise{First of all, the BN-based approaches, such as \break \textsf{BayesCard}, are generally interpretable and predictable, thus easy to debug for DBMS analytics.}
More importantly, a \CE method friendly for system deployment should have faster training time and lightweight model size and \textsf{BayesCard} has the dominant advantage over the other ML-based data-driven methods in these two aspects because of its underlying Bayesian model.
Specifically, from both training time and model size aspects, these methods can be ranked as \textsf{BayesCard} $<<$ \textsf{FLAT}/\textsf{DeepDB} $<$ \revise{\textsf{NeuroCard}$^E$}. We provide the detailed reasoning as follows.

\textsf{BayesCard} proposes an accelerated model construction process of BN using probabilistic programming. Its model training time is roughly 100 times faster than the other three data-driven methods. Moreover, the BNs in \textsf{BayesCard}, which utilize the attribute conditional independence to reduce the model redundancy, are naturally compact and lightweight.

\textsf{FLAT} and \textsf{DeepDB} recursively learn the underlying FSPN and SPN models. 
Their training time is less stable and varies greatly with the number of highly correlated attributes in the datasets. 
Thus, we observe a much longer training time on STATS than on the IMDB dataset for these two methods.
The SPNs in \textsf{DeepDB} iteratively split the datasets into small regions, aiming to find local independence between attributes within each region. 
However, in presence of highly correlated attributes (e.g. STATS), the SPNs tend to generate a long chain of dataset splitting operation, leading to long training time and a very large model size. The FSPNs in \textsf{FLAT} effectively address this drawback of SPNs by introducing the factorize operation but their training time and model size suffer greatly for datasets with a large number of attributes (e.g. STATS) because of the recursive factorize operations.

\revise{The training of \textsf{NeuroCard}$^E$ is particularly long and its size is also the largest on STATS because its join schema does not form a tree. As mentioned in Section~\ref{sec: plan-algo}, the original \textsf{NeuroCard} only supports tree-structured schemas. 
Thus, \textsf{NeuroCard}$^E$ extracts $16$ tree sub-structures from STATS schema graph and train one model for each tree. Therefore, we argue that extending \textsf{NeuroCard} for non-tree-structured schemas can greatly improve its practicality.}

\subsection{Model Update}
\label{sect6.3}
Model update is a crucial aspect when deploying a \CE method in OLTP databases. Frequent data updates in these DBs require the underlying \CE method to swiftly update itself and adjust to the new data accurately. In the following, we first provide our observation regarding the updatability of ML-based query-driven methods and then provide the update experimental settings and results for ML-based data-driven methods on the STATS dataset.

\smallskip
\noindent \textbf{O9: Existing query-driven \CE methods are impractical for dynamic DBs.} The query-driven models require a large amount of executed queries to train their model, which might be unavailable for a new DB and very time-consuming to generate (e.g. 146 STATS-CEB queries take more than ten hours to execute). More importantly, they need to recollect and execute the queries whenever datasets change or query workload shifts. Therefore, they can not keep up with the frequent data update in dynamic DBs.

\smallskip
\noindent \underline{\textbf{Experimental Settings:}} To simulate a practical dynamic environment, we split the STATS data into two parts based on the timestamps of tuples. We first train a stale model for each method on the data created before 2014 (roughly $50\%$) and insert the rest of the data to update these models. We only test the data insertion scenario since some methods (\revise{\textsf{NeuroCard}$^E$} and \textsf{DeepDB}) do not support data deletion. We use the update algorithm in these methods' publicly available source code. 

\begin{table}
	\caption{Update performance of \CE algorithms.}
	\vspace{-1em}
	\label{tab:update}
	\scalebox{0.8}{
		\begin{tabular}{@{}ccccc@{}}
			\toprule
			\rowcolor{mygrey}
			\bf Criteria & \bf \revise{\textsf{NeuroCard}$^E$} & \bf \textsf{BayesCard} & \bf \textsf{DeepDB} & \bf \textsf{FLAT} \\ \cmidrule{1-5}
			Update time & 5,569s & \textbf{12s} & 248s  & 360s \\
			Original E2E time (Table~\ref{tab:overall}) &11.85h &7.16h &6.46h &5.80h \\
			E2E time after update & 13.94h & 7.16h & 6.72h & 7.04h \\ \cmidrule{1-5}
		\end{tabular}
	}
	\vspace{-2em}
\end{table}

\noindent \underline{\textbf{Experimental Results:}} 
As shown in Table~\ref{tab:update}, we record the time these methods take to update their stale models and evaluate the end-to-end query performance of the updated models on STATS-CEB queries. We also cite the original model performance from Table~\ref{tab:overall} as comparison baselines. We first summarize the most important observation based on this table and then provide detailed reasoning from the update speed and accuracy perspectives.

\smallskip
\noindent \textbf{O10: Data-driven \CE methods have the potential to keep up with fast data update and can be applied in dynamic DBs.} Specifically, \textsf{BayesCard} takes $12s$ to update itself for an insertion of millions of tuples on multiple tables in the DB. More important, its end-to-end performance is unaffected by this massive update, thus very suitable for dynamic DBs.  

\textbf{Update speed} of these methods can be ranked as \textsf{BayesCard} $>>$ \textsf{DeepDB} $>$ \textsf{FLAT} $>$ \revise{\textsf{NeuroCard}$^E$}. \textsf{BayesCard} preserves its underlying BN's structure and only incrementally updates the model parameters. Since its model size is relatively small, the update speed is more than $20$ times faster than others. \textsf{DeepDB} and \textsf{FLAT} also preserves their underlying structure of SPN and FSPN but as their structures are significantly larger, the incrementally updating their model parameters still take a large amount of time.

\textbf{Update accuracy} can be ranked as \textsf{BayesCard} $>$ \textsf{DeepDB} $>$ \textsf{FLAT} $>$ \revise{\textsf{NeuroCard}$^E$}. \textsf{BayesCard}'s underlying BN's structure captures the inherent causality, which is unlikely to change when data changes. Therefore, \textsf{BayesCard} can preserve its original accuracy after model update (i.e. same as its comparison baseline). The structures of SPN in \textsf{DeepDB} and FSPN in \textsf{FLAT} are learned to fit the data before the update and cannot extrapolate well to the newly inserted data. Therefore, only updating the model parameters will cause modeling inaccuracy (i.e. we observe a drop in their end-to-end performance when compared with their baselines).

\begin{table*}
	\caption{Comparison between Q-Error and P-Error.}
	\vspace{-1em}
	\resizebox{0.9\textwidth}{!}{
		\begin{tabular}{cc|ccc|ccc||cc|ccc|ccc}
			\toprule
			\rowcolor{mygrey}
			& \multicolumn{6}{c}{\bf JOB-LIGHT Workload} & & &
			\multicolumn{6}{c}{\bf STATS-CEB Workload} & \\
			\cmidrule{1-8} \cmidrule{9-16}
			\rowcolor{mygrey}
			\multirow{2}{*}{\bf Method} & {\bf Execution Time} & \multicolumn{3}{c}{\bf Q-Error}& & \bf P-Error & &
			\multirow{2}{*}{\bf Method} &
			{\bf Execution Time} & 
			\multicolumn{3}{c}{\bf Q-Error}& \multicolumn{3}{c}{\bf P-Error}
			\\ \cmidrule{3-8} \cmidrule{11-16}
			\rowcolor{mygrey}
			& {\bf (Descending Order)} & \bf 50\% & \bf 90\% & \bf 99\% & \bf 50\% & \bf 90\% & \bf 99\% & & {\bf (Descending Order)} & \bf 50\% & \bf 90\% & \bf 99\% & \bf 50\% & \bf 90\% & \bf 99\%
			
			\\ \hline
			
			\revise{\textsf{UniSample}} &$>25$h & $5.754$ & $466.6$ & $3\cdot 10^5$ & $1.139$ & $4.651$ & $536.3$ & \revise{\textsf{UniSample}} & $4.84$h & $3.259$&$135.4$ &$6\cdot 10^4$ & $1.000$ & $1.345$ & $3.593$
			\\
			
			\textsf{LW-XGB}  &$4.31$h& $5.303$ & $261.2$ &  $2\cdot 10^5$ & $1.0$ & $3.728$& $15.48$ & \textsf{LW-XGB}   &$>25$h&$5.652$&$453.2$& $3\cdot 10^5$ &$1.742$& $6.627$ & $526.2$
			\\
			
			\revise{\textsf{WJSample}} & \revise{$4.15$h} & \revise{$2.445$}   & \revise{$114.8$}  & \revise{$4 \cdot 10^3$} & \revise{$1.000$} & \revise{$1.223$} &  \revise{$3.127$} & \revise{\textsf{WJSample}} & \revise{$19.86$h} & \revise{$3.452$} & \revise{$351.4$} & \revise{$8 \cdot 10^4$} & \revise{$1.103$} & \revise{$4.954$} & \revise{$501.6$} \\
			
			\revise{\textsf{UAE}} & \revise{$3.71$h} &\revise{$1.167$} &\revise{$2.395$}&\revise{$17.196$} & \revise{$1.145$} & \revise{$2.983$} & \revise{$39.58$} & \textsf{NeuroCard} &$11.85$h& $951.4$& $9\cdot 10^5$ & $6\cdot 10^8$& $1.193$ & $2.844$ & $1\cdot 10^3 $\\
			
			\textbf{PostgreSQL} & \textbf{3.67h} & $\mathbf{1.810}$ & $\mathbf{8.089}$ & $\mathbf{51.80}$ & $\mathbf{1.000}$ & $\mathbf{1.408}$ & $\mathbf{2.329}$ & \revise{\textsf{UAE}} & \revise{$11.46$h}  & \revise{$3.239$}& \revise{$130.8$}&  \revise{$1.1\cdot 10^4$} &  \revise{$1.187$} &  \revise{$2.420$} &  \revise{$7.873$} \\
			
			\revise{\textsf{UAE-Q}} & \revise{$3.65$h} &  \revise{$1.112$} &  \revise{$2.760$} &  \revise{$308.1011$} & \revise{$1.034$} & \revise{$2.356$} & \revise{$9.958$} & \textbf{PostgreSQL} &\textbf{11.34h} & $\mathbf{1.439}$ & $\mathbf{11.08}$ & $\mathbf{2\cdot 10^3}$  & $\mathbf{1.000}$ & $\mathbf{1.595}$ & $\mathbf{12.35}$ \\ 
			
			
			\textsf{LW-NN} & $3.63$h& $5.713$ & $112.9$  & $4\cdot 10^4$ & $1.104$ & $2.448$& $43.67$ & 
			\textsf{LW-NN}  &$11.33$h  & $15.73$& $832.9$ & $2\cdot 10^4$ &$1.159$ & $4.651$  & $18.02$\\
			
			\textsf{MSCN} &\revise{$3.50$h}&  \revise{$1.724$} & \revise{$12.68$} & \revise{$182.4$}& \revise{$1.126$}& \revise{$2.805$}& \revise{$9.747$} & 
			\revise{\textsf{UAE-Q}} & \revise{$11.03$h} & \revise{$2.875$} & \revise{$20.613$} & \revise{$1\cdot 10^4 $}  & \revise{$1.145$} & \revise{$4.001$} & \revise{$13.14$}\\
			
			\textsf{NeuroCard} &$3.29$h& $2.153$& $10.17$ & $26.89$ & $1.011$ & $1.984$ & $2.747$ & \textsf{MSCN}  &\revise{$8.11$h}  & \revise{$20.92$} & \revise{$392.3$} & \revise{$1\cdot 10^4$} & \revise{$1.138$} &\revise{$4.031$} & \revise{$11.11$}\\
			
			\textsf{DeepDB}  &$3.28$h&$1.332$&$2.257$& $33.29$ & $1.000$ &  $1.061$ & $1.528$ & \textsf{BayesCard} &$7.16$h  &$1.182$ & $50.40$ & $156.4$ & $1.000$& $1.582$& $6.843$\\
			
			\textsf{FLAT} &$3.21$h& $1.230$& $1.987$  &  $15.71$ & $1.000$ & $1.062$ & $1.528$ & \textsf{DeepDB} &$6.46$h & $2.451$ & $22.37$ & $1\cdot 10^3$ & $1.030$ & $1.833$ & $6.819$ \\
			\textsf{BayesCard} &$3.18$h & $1.145$ & $2.835$ & $10.33$& $1.000$ & $1.002$ & $1.123$ & \textsf{FLAT} &$5.80$h & $1.675$ & $10.44$ &$768.8$ & $1.000$ &$1.346$ &$5.546$ \\
			\hline
	\end{tabular}}
	\vspace{-1em}
	\label{tab:metric}
\end{table*}

\section{Is Current Metric Good Enough?}
\label{sec:analysis}

\revise{
Most of the existing works~\cite{zhu2020flat,yang2020neurocard,hilprecht2019deepdb,kipf2018learned,dutt2019selectivity} use Q-Error~\cite{moerkotte2009preventing} to evaluate the quality of their \CE methods. However, the ultimate goal of \CE is to generate query plans with faster execution time. Therefore, we explore whether Q-Error is a good metric to fulfill this goal in this section. We first analyze the correlations between Q-Error and query execution time in Section~\ref{sect7.1}. The results show that smaller Q-Error does not necessarily lead to shorter execution time. Thus, we identify the limitations of Q-Error and propose another metric called P-Error in Section~\ref{sect7.2}.
We show that P-Error has better correspondence to query execution time and advocate it to be a potential substitution of Q-Error.
}

\subsection{Problems with Q-Error}
\label{sect7.1}

Q-Error is a well-known metric to evaluate the quality of different \CE methods.
It measures the relative multiplicative error of the estimated cardinality from the actual one as:
\begin{equation*}
\textbf{Q-Error} =  \max(\frac{\text{Estimated Cardinality}}{\text{True Cardinality}}, \frac{\text{True Cardinality}}{\text{Estimated Cardinality}}).
\end{equation*}
\revise{
Q-Error penalizes both overestimation and underestimation of the true cardinality. However, existing works have not investigated whether Q-Error is good evaluation metric for \CEend. I.e, would \CE methods with smaller Q-Errors definitely generate query plans with shorter execution time and vice versa? To answer this question, we revisit the experimental results. Table~\ref{tab:metric} reports the distributions ($50\%, 90\%  \text{ and } 99\%$ percentiles) of all sub-plan queries' Q-Errors generated by different \CE methods on both JOB-LIGHT and STATS-CEB benchmarks. We sort all \CE methods in a descending order of their execution time. 
From a first glance, we derive the following observation:
}

\noindent \textbf{O11: The Q-Error metric can not serve as a good indicator for query execution performance.} 
This observation is supported by a large amount of evidence from Table~\ref{tab:metric}. We list three typical examples on STATS-CEB as follows: 
\revise{
1) \textsf{NeuroCard}$^E$ has the worst Q-Errors in all methods, but its execution time is comparable to \textsf{PostgreSQL} and much better than histogram and sampling based methods and \textsf{LW-XGB};
2) \textsf{BayesCard} has the best Q-Errors, yet execution time is $1.4h$ slower than \textsf{FLAT};
and 3) the Q-Errors of \textsf{MSCN} are significantly worse than \textsf{PostgreSQL}, but the execution time of \textsf{MSCN} largely outperforms it.
}

\revise{
Next, we analyze the underlying reasons behind $O11$. This is particularly important as the DB communities have made great efforts in purely optimizing the Q-Error of \CE methods, but sometimes neglect the ultimate goal of \CE in DBMS. As shown in Section~\ref{sec: plan-sys}, the \CE method would be invoked for multiple sub-plan queries to decide the query plan. The estimation errors of different sub-plan query have different impact on the final query plan performance. However, the Q-Error metric could not distinguish this difference and regard the estimation errors of all queries equally. This would cause the phenomenon that a more accurate estimation measured by Q-Error may lead to a worse query execution plan in reality. We list two typical scenarios in the benchmark where Q-Error fails to distinguish the difference as follow:
}

\noindent \textbf{\revise{O12: Q-Error does not distinguish queries with small and large cardinality that have the same Q-Error value but matter differently to the query plan.}} 
\revise{
For Q-Error, an estimation $1$ for true cardinality of $10$ has the same Q-Error as an estimation $10^{11}$ for true cardinality $10^{12}$. 
The previous case may barely affect the overall query plan, whereas the latter one can be catastrophic since the (sub-plan) queries with large cardinalities dominate the overall effectiveness of the query plan (shown in O5). For example, in Figure~\ref{fig:q57plans}, the overall Q-Error of \textsf{BayesCard} over all sub-plan queries of Q57 is better than \textsf{FLAT}. However, only for the root query which matters most importantly to the query execution time, \textsf{BayesCard} fails to correctly estimate and leads to a much slower plan. 
}

\noindent \textbf{\revise{O13: Q-Error can not distinguish between query underestimation and overestimation that have the same Q-Error value but matter differently to the query plan.}}
\revise{
For Q-Error, an underestimation $10^{9}$ for true cardinality $10^{10}$ is the same as an overestimation of $10^{11}$. These two estimations are very likely to lead to different plans with drastically different execution time. Recall the Q57 example, \textsf{BayesCard} underestimates the cardinality of the root query by $7\times$ and selects a ``merge join'' operation. We test this query but injecting a $7\times$ overestimation for this sub-plan query, and it then selects the ``hash join'' operation with twice faster time.
}


\revise{
As a result, the Q-Error metric does not consider the importance of different sub-plan queries and may mislead the query plan generation, so it is not a good optimization goal for \CE methods.
}

\subsection{An Alternative Metric: P-Error }
\label{sect7.2}

Obviously, the best way to evaluate the quality of a \CE method is to directly record its query execution time on some benchmark datasets and query workloads (e.g. JOB-LIGHT and STATS-CEB).
\revise{
However, this is time consuming and not suitable for the situations where fast evaluation is needed, e.g., hyper-parameter tuning.
A desirable metric should be fast to compute and  simultaneously correlated with the query execution time. In the following, we propose the P-Error metric to fulfill this goal and quantitatively demonstrate that P-Error can be a possible substitute for Q-Error.
}

\noindent \underline{\textbf{P-Error metric for \CE:}}
\revise{
Although obtaining the actual query execution time is expensive, we could 
approximate it using the built-in component in DBMS. Note that, given a query plan, the cost model of a DBMS could output an estimated cost, which is designed to directly reflect the actual execution time. Inspired by the recent research~\cite{negi2021flow}, we believe that the estimated cost could serve as a good metric for evaluating the \CE methods.}

\revise{
Specifically, given a query $Q$ and a \CE method $A$, let $C^{\text{T}}$ and $C^{\text{E}}$ denote the set of true and estimated cardinality of all sub-plan queries of $Q$. 
When $C^{\text{E}}$ is fed into the query optimizer, it would generate a query plan $P(C^{\text{E}})$ of $Q$.
During the actual execution of this query plan, the true cardinalities of all sub-plan queries along this plan will be instantiated.
Therefore, to estimate the execution cost of $P(C^{\text{E}})$, we inject the true cardinality of sub-plan queries $C^{\text{T}}$ into the DBMS. The DBMS will output an estimated cost based on this query plan, which is highly correlated to the actual time for an accurate cost model. Following prior work~\cite{negi2021flow}, we choose PostgreSQL to calculate this estimated cost, which is denoted as $PPC(P(C^{\text{E}}), C^{\text{T}})$.
}

\revise{
Ideally, if the cost model is accurate, the query plan $P(C^{\text{T}})$ found by the true cardinality $C^{\text{T}}$ should be optimal, i.e. $PPC(P(C^{\text{T}}), C^{\text{T}}) = \min_{C} PPC(P(C), C^{\text{T}})$. Therefore, we define
\begin{equation*}
    \textbf{P-Error} =  PPC(P(C^{\text{E}}), C^{\text{T}})/ PPC(P(C^{\text{T}}), C^{\text{T}})
\end{equation*}}

\vspace{-1em}

\noindent \revise{as our \CE metric. The P-Error for an existing workload of queries can be computed instantaneously using the modified
plugin \textsf{pg\_hint\_plan} provided in~\cite{pghintplan} as long as we pre-compute and store the true cardinalities of all sub-plan queries.}

\revise{
In P-Error, the effectiveness of a \CE method's estimation $C^{\text{E}}$ is measured on the plan cost level. The impact on the estimation error of each sub-plan query is reflected by its importance in generating the query plan $P(C^{\text{E}})$ (e.g. small or large cardinality, underestimation or overestimation, etc.). 
}

\revise{
Notice that, in real-world DBMS, the cost model can sometimes be inaccurate~\cite{leis2015good}, which may lead to worse query plans with better estimated cost, i.e., $PPC(P(C^{\text{T}}), C^{\text{T}})$ may be not the minimal cost over all query plans. However, this is not an issue as $PPC(P(C^{\text{T}}), C^{\text{T}})$ is identical to different \CE methods, we could always compare their relative performance using P-Error no matter $P(C^{\text{T}})$ is optimal or not. 
Meanwhile, we find that $P(C^{\text{T}})$ is optimal in most cases.
On our STATS benchmark, the query plan generated by the true cardinality is optimal on more than $98\%$ queries using the default cost model of PostgreSQL. 
}

\myskip
\noindent
\revise{
\underline{\textbf{Advantages of P-Error Metric:}}
In Table~\ref{tab:metric}, we report the P-Error distributions ($50\%, 90\%, 99\%$ percentiles) over the query workload of all \CE methods and derive the following observation:
}

\noindent \textbf{O14: P-Error is more highly correlated to the query execution time than Q-Error.} 
\revise{
We can roughly see that methods with better runtime tend to have smaller P-Error (e.g. \textsf{FLAT} has the best P-Error). We also compute the correlation coefficients between the query execution time and Q-Error/P-Error. On the STATS-CEB query workload, the value between $50\%$ and $90\%$ percentiles of Q-Error distribution w.r.t.~query time is $0.036$ and $0.037$. Whereas, the value between $50\%$ and $90\%$ percentiles of P-Error distribution w.r.t.~query time is $0.810$ and $0.838$. This indicates that P-Error is a better correspondence to the query execution time than Q-Error. 
}

\revise{
In addition, P-Error is more convenient as it outputs a single value on the plan cost level whereas Q-Error outputs a value for each sub-plan query of $Q$. Therefore, P-Error makes an attempt to overcome the limitations of Q-Error and is shown to be more suitable to measure the actual performance of \CE methods.
}



\section{Conclusions and Future Work}
In this paper, we establish a new benchmark for \CE, which contains the complex real-world dataset STATS and the diverse query workload STATS-CEB. This new benchmark helps to clearly identify the pros and cons of different \CE methods. In addition, we propose the new metric P-Error as a potential substitute for the well-known Q-Error. \revise{Based on the exhaustive experimental analysis, we derive a series of important observations that will provide the DBMS community with a holistic view of the \CE problem and help researchers design more effective and efficient methods. We summarize the following key take-away messages:}

\revise{
$\bullet$ \textbf{Overall performance of \CE methods:} 
Both estimation accuracy and inference time matters to the end-to-end performance of query optimization. Some ML-based data-driven \CE methods built on top of PGMs, such as \textsf{FLAT}, \textsf{DeepDB} and \textsf{BayesCard}, can largely improve the end-to-end performance as they make the right balance in tuning the strictness of independence assumption to keep inference efficiency and model effectiveness. Whereas, the ML-based query-driven methods barely have any improvement over the \textsf{PostgreSQL} baseline.
}

\revise{
$\bullet$ \textbf{Practicality aspects of \CE methods:}
Some traditional methods and ML-based data-driven \CE methods, such as \textsf{PessEst} and \textsf{BayesCard}, can be applied in real-world dynamic DBMS as their updating speed could keep track of the fast data update pace. Whereas, existing ML-based query-driven methods are inherently impractical for DBs with frequent data updates.
}

\revise{
$\bullet$ \textbf{Challenges for multi-table join queries:} 
Learning one large data-driven model on the (sample of) full outer join of all tables has poor scalability and low accuracy for numerous tables in a DB. However, ML-based data-driven \CE methods exhibit degrading performance for queries with an increasing number of join tables. This brings big challenges for the actual deployment of ML-based methods in DBMS.}

\revise{
$\bullet$ \textbf{Importance of different queries:} 
The estimation errors of different sub-plan queries matter differently to the final query plan quality. Sometimes, accurate estimation of queries with large cardinalities is much important than the small ones. Therefore, optimizing the accuracy of \CE methods along the Q-Error metric does not always generate high quality query plans.
}

\smallskip
\revise{
Based on the takeaway messages, we point out the following future research directions (RD) for \CE methods:}

\indent $\bullet$ \textbf{RD1}: 
Enlarging the application scope of ML-based \CE methods to support complex queries with ``LIKE'' predicates and cyclic joins.

\indent 
\revise{
$\bullet$ \textbf{RD2}: 
Combining different models together to design \CE methods that could adjust the estimation accuracy and inference cost to fit different settings, e.g., OLAP and OLTP workload.
}

\indent 
\revise{
$\bullet$ \textbf{RD3}: 
Optimizing \CE methods towards the end-to-end performance, i.e., using our new P-Error metric as the objective function and fine-tuning the estimation quality on important, possible large, sub-plan queries.
}

\indent 
\revise{
$\bullet$ \textbf{RD4}: 
Designing new factorization methods for join queries to better balance the estimation accuracy and training/inference cost.
}

\clearpage

\bibliographystyle{ACM-Reference-Format}
\bibliography{main}

\end{document}